\documentclass[12pt]{article}

\usepackage{amssymb}
\usepackage{amsmath}
\begin{document}
\title{Nonlinear Gravitational Lagrangians revisited}
\author{Guido Magnano\\ \small Universit\`a di Torino, Dipartimento di Matematica ``G.~Peano'' \\ \small via Carlo Alberto 10, I-10123 Torino, Italy\\ \small guido.magnano@unito.it }
\date{December 2015}
\maketitle

\begin{abstract}
\noindent The Legendre transformation method, applied in 1987 to deal with purely metric gravitational Lagrangians with nonlinear dependence on the Ricci tensor, is extended to metric--affine models and is shown to provide a concise and insightful comparison of the dynamical content of the two variational frameworks.
\end{abstract}

\section{Thirty years after}

In june 1987, the article \emph{Nonlinear Gravitational Lagrangians}\cite{NGL} was published in \emph{General Relativity and Gravitation}. The core of the article was the application of a (generalized) Legendre transformation to models of gravity in which the Lagrangian depended in a nonlinear way on the Ricci tensor of the space-time metric. The method turned out to work for both Lagrangians depending only on the scalar curvature and Lagrangians depending on the full Ricci tensor, although the appropriate Legendre transformation is different in the two cases. A parallel, independent work by A.~Jakubiec and J.~Kijowski discussing the application of the Legendre transformation to gravitational theories \cite{JK} was published in the same volume of the GRG journal. Both papers followed an earlier proposal by Kijowski \cite{K79}; a mathematical framework for the Legendre transformation in higher--order field theories, based on Poincar\'e-Cartan forms rather that on symplectic geometry, was introduced in \cite{MFF90}.

The common reception of these results was that any nonlinear gravitational Lagrangian is equivalent to the well-known Einstein-Hilbert Lagrangian (with additional, minimally coupled terms for some auxiliary field variable) \emph{up to a redefinition of the metric.} Whenever the Lagrangian depends only on the scalar curvature (which is universally referred to as the $f(R)$ case), the redefinition of the metric consists in a conformal rescaling: this was in line with previous observations made by different authors (for a comprehensive list of early references, see \cite{MS1}), which led most researchers, in the subsequent years, to regard the Legendre transformation and the metric redefinition as a single operation that can be performed to exhibit a dynamical equivalence between ``nonlinear gravity'' and General Relativity. In more recent years, this has been repeatedly rephrased as the existence of two ``frames'': the \emph{Jordan frame}, defined by the original metric obeying a fourth--order field equation, and the \emph{Einstein frame} defined by the new metric and leading to the familiar Einstein equation. 

Many authors have then addressed the question whether the ``physical frame'' should be assumed to be the Jordan or the Einstein one. Actually, the question had previously arisen for the Jordan-Brans-Dicke scalar-tensor theory, whereby a conformal rescaling was known to recast the Lagrangian in the usual Einstein-Hilbert form: this is the origin of the widespread ``Jordan/Einstein frame'' terminology (although the reference to a ``frame'' is potentially misleading: the two pictures, in fact, are not related by a change of coordinates or of reference frame). The analogous debate for the $f(R)$ models was started by Carl Brans in a letter \cite{Brans} in response to \cite{NGL}: Brans pointed out that, after redefining the metric, the gravitational mass of bodies becomes dependent on the point in space and time, and stress--energy conservation laws are broken. In reply to \cite{Brans}, it was observed that assuming minimal coupling between matter fields and the original metric, in the construction of a gravity model, entails that one has already postulated that the original metric is the ``physical'' one: in that case, indeed, one cannot maintain that the new metric is physical as well, and it is not surprising that the interaction of the new metric with matter becomes somehow ``unphysical''. But the argument can be reversed (stress--energy conservation, in particular, holds in Einstein frame if the covariant derivatives are taken with respect to the new metric) and therefore it cannot be invoked to claim that either metric is unphysical \cite{rBrans}. 

In other terms, the question of which tensor field has to be identified with the physical spacetime metric should be addressed \emph{before} devising the additional terms which describe the gravitational interaction of matter fields. The equivalence of $f(R)$ models with General Relativity plus a scalar field, holding in vacuum, is indeed broken if matter fields are coupled to the original metric tensor as if the latter were the physical one. 

A criterion to decide which metric is the physical one, based on structural properties on the model (positivity of energy), has been proposed in \cite{MS1}, and we shall recall it below; but this has not settled the dispute, seemingly, even in cases where the criterion would be applicable. In fact, the attitude of the various authors depends on their purposes for introducing a $f(R)$ gravity model. Some authors regarded such models as a nice way to make a ``quintessential'' scalar field appear not as an additional field, but rather as an additional degree of freedom which is already contained in the original metric and reveals itself as a separate field upon switching to the Einstein frame. In that case, the original (Jordan--frame) metric is regarded as a sort of \emph{unifying} field variable, while the role of physical metric is attributed to the Einstein--frame metric. In the last decade, the prevailing viewpoint seems to be the opposite one: many authors explore the cosmological effects of adding ordinary cosmic matter terms to $f(R)$ Lagrangians, without considering any metric redefinition. Meanwhile, most authors also turned to what appears to be a deep change in the basic assumptions of the model: namely, the affine connection $\Gamma$ whose curvature enters the Lagrangian is no longer assumed to be the Levi--Civita connection of the metric $g$, but instead becomes an independent field. This is the metric--affine, or Palatini, version of the action principle \cite{Cap}. 

Even in the Palatini setup it is found that, upon suitable assumptions, a metric redefinition (which in the vacuum reduces to the identity) leads to a dynamically equivalent picture where the dynamical connection \emph{is} the Levi--Civita connection of the new metric, which in turn obeys Einstein equations; however, it is exactly the possibility of coupling cosmic matter to a metric describing a \emph{different} spacetime geometry (with respect to the connection) which seems appealing, in order to fit the observational data. The fact that $\Gamma$ coincides with the Levi--Civita connection for a conformally-rescaled metric (as a consequence of the field equations), on the other hand, ensures that light propagation under the original metric $g$ is physically compatible (in the sense of Ehlers--Pirani--Schild \cite{EPS}) with the free fall of massive bodies described by the geodesics of $\Gamma$.

The Palatini versions of $f(R)$ models are now often referred to as ``extended gravity theories'' \cite{Cap2}. Although a detailed comparison between the equations of Palatini f(R) models, purely metric f(R) models and scalar-tensor models has been presented by various authors, the involvement of field redefinitions tends to obscure the exact extent of the differences between the dynamical contents of these models. 

Palatini $f(R)$ models were not considered in \cite{NGL}, and one may suspect that the approach of that article is intrinsically confined to the purely metric setup. On the contrary, in the sequel we show that the (distinct) ideas of Legendre transformation and of metric redefinition, both introduced in \cite{NGL}, provide a nice way to
\begin{description}
\item[(i)] obtain a systematic comparison between Palatini and purely metric models; such a comparison, through a suitable scalar-tensor reformulation of both models, has been already discussed in the previous literature (see e.g.~\cite{Cap}), but our approach reveals in a more direct way that the difference between the purely metric variation and the Palatini variation of a given $f(R)$ Lagrangian merely amounts to swapping (with a relevant change of sign) the dynamical term of a scalar field from the ``Einstein frame'' to the ``Jordan frame'' Lagrangian;
\item[(ii)]  generalize this result to Lagrangians depending on the (symmetric part) of the Ricci tensor; the Palatini versions of such models have been studied by some authors (\cite{Bor},\cite{LBM}), but an insightful comparison with their purely metric counterparts is still lacking.
\end{description}

The following discussion is thus aimed at exploiting the Legendre transformation to put the Palatini and the purely metric gravitational models on the same footing, so that the actual differences between them can be clearly seen already at the level of the action principle. 

\section{Purely metric theories}

Let $g_{\mu\nu}$ be a metric (of the appropriate signature) on the spacetime manifold $M$ (we shall assume that $\dim(M)=4$, but this is not crucial for most of the subsequent discussion). We shall denote by $\left\lbrace{}^\lambda_{\mu\nu}\right\rbrace_g$ the Christoffel symbols, i.e.~the components of the Levi--Civita connection, for the metric $g_{\mu\nu}$; the scalar density $\sqrt{|g|}$ will be the component of the associate volume form; $R^{\alpha}_{\mu\beta\nu} $, $R_{\mu\nu}\equiv R^{\alpha}_{\mu\alpha\nu}$ and $R\equiv R_{\mu\nu}g^{\mu\nu}$ will denote, respectively, the Riemann tensor, the Ricci tensor and the curvature scalar of the metric. 

We recall that a covariant Lagrangian cannot depend only on first derivatives of the metric, unless a separate affine connection is used (either as a fixed background or as an independent field variable): thus, in purely metric theories the gravitational Lagrangian should depend at least on second derivatives of the metric, and only through the components of the Riemann tensor. To the vacuum Lagrangian one adds interaction terms for the other physical fields (that we collectively denote by $\Phi$), possibly depending on first derivatives of the metric. In the sequel, we shall use the jet notation $j^k x$ to indicate that a function depends on some field $x$ and on its spacetime derivatives up to the $k$-th order. 

Although Lagrangians depending on the full Riemann tensor have sometimes been considered in the literature, here we shall focus on models depending only on the Ricci tensor, as was done in \cite{NGL}:
\begin{equation}
L=f(R_{\mu\nu},g_{\mu\nu})\sqrt{|g|}+L_{\mathrm{mat}}(j^1 \Phi,j^1 g).\label{HOG}
\end{equation}

A particular case is 
\begin{equation}
L=f(R)\sqrt{|g|}+L_{\mathrm{mat}}(j^1 \Phi,j^1 g)\label{NL},
\end{equation}
whereby the Lagrangian depends solely on the curvature scalar. The Einstein--Hilbert Lagrangian belongs to this class, with $f(R)\equiv R$, and is the only one which generates second--order Euler-Lagrange equations for the metric: for any nonlinear function $f(R)$ (or, more generally, $f(R_{\mu\nu},g_{\mu\nu})$) the equations for the gravitational metric will be of order four. The function $f$ is generally assumed to be differentiable, but a class of models have been proposed in recent years \cite{CCT} where $f$ has a pole at $R=0$.

Variation of (\ref{HOG}) with respect to the metric produces the equation
\begin{equation}
J^{\alpha\beta}+\frac{1}{2}\left(P^{\beta\nu}{}_{;\mu\nu}g^{\alpha\mu}+P^{\alpha\nu}{}_{;\mu\nu}g^{\beta\mu}-P^{\mu\nu}{}_{;\mu\nu}g^{\alpha\beta}-P^{\alpha\beta}{}_{;\mu\nu}g^{\mu\nu}\right)=\kappa T^{\alpha\beta}\sqrt{|g|}\label{HOGeq}
\end{equation}
where
\begin{eqnarray*}
P^{\alpha\beta} &\equiv& \frac{\partial f}{\partial R_{\alpha\beta}}\sqrt{|g|}, \\
J^{\alpha\beta} &\equiv& \left(\frac{\partial f}{\partial g_{\alpha\beta}}+\frac{1}{2}g^{\alpha\beta}f\right)\sqrt{|g|},
\end{eqnarray*}
$T^{\alpha\beta}\sqrt{|g|}$ is (up to a constant factor) the variational derivative of the scalar density $L_{\mathrm{mat}}$ with respect to the metric and the semicolon denotes covariant differentiation under the connection $\left\lbrace{}^\lambda_{\mu\nu}\right\rbrace_g$.

For the Lagrangian (\ref{NL}), equation (\ref{HOGeq}) reduces to 
\begin{equation}
f'(R)R^{\alpha\beta}-\frac{1}{2}f(R)g^{\alpha\beta}+f'(R)_{;\mu\nu}\left(g^{\mu\nu}g^{\alpha\beta}-g^{\mu\alpha}g^{\nu\beta}\right)=\kappa T^{\alpha\beta},\label{NLeq}
\end{equation}
where $f'(R)\equiv\frac{df}{dR}$.

The idea of performing a Legendre transformation of $L$ with respect to the Ricci tensor is due to J.~Kijowski \cite{K79}: he proposed that, in a gravity theory, the fields which jointly describe gravity (i.e. the metric, defining infinitesimal spacetime separation, and the affine connection, defining the free--fall worldlines and the inertial reference frames) should be regarded as \emph{mutually conjugate variables} in the sense of symplectic geometry. More precisely, the metric is conjugate to the connection, whereby the components of the (symmetrized) Ricci tensor of the connection play the role of the ``velocity components'' in the Legendre transformation. 

Originally, this idea was exploited to introduce a metric tensor in purely affine models, whereby the Lagrangian contains only an affine connection $\Gamma$ (e.g. in the Einstein-Eddington model, see \cite{FK} and references therein): a tensor density $\pi^{\mu\nu}$ is then introduced through the Legendre map
\begin{equation}
\pi^{\mu\nu}=\frac{\partial L}{\partial R_{(\mu\nu)}}
\end{equation}
and subsequently converted into a symmetric tensorfield through multiplying $\pi^{\mu\nu}$ by the square root of (the absolute value of) its determinant. For relevant Lagrangians the resulting tensorfield is generically nondegererate, and the appropriate signature can be imposed, for instance, on the Cauchy data.

In metric-affine models and purely metric models, a metric $g_{\mu\nu}$ is already present among the dynamical variables: not only there is no apparent reason to introduce a new field conjugate to the connection, but one should expect that such a new field would \emph{not} coincide with $g_{\mu\nu}$ (the only exception being the Einstein-Hilbert Lagrangian, as is easy to check). In spite of that, for the Lagrangian (\ref{HOG}) we do introduce the new tensorfield and call it $\tilde{g}^{\mu\nu}$:
\begin{equation}
\tilde{g}^{\mu\nu}\sqrt{|\tilde{g}|}=\frac{\partial L}{\partial R_{\mu\nu}}=\frac{\partial f}{\partial R_{\mu\nu}}\sqrt{|g|}.
\label{LM1}
\end{equation}
(the covariant components $\tilde{g}_{\mu\nu}$ are defined as the entries of the inverse matrix, $\tilde{g}_{\alpha\nu}\tilde{g}^{\nu\beta}\equiv\delta^{\beta}_{\alpha}$, \emph{not} by lowering the indices with the original metric). In order to perform a Legendre transformation, however, the map (\ref{LM1}) is useful only if it can be inverted, i.e.~if a tensor function $r_{\mu\nu}(g_{\mu\nu},\tilde{g}_{\mu\nu})$ exists such that
\begin{equation}
\sqrt{|g|}\left.\frac{\partial f}{\partial R_{\mu\nu}}\right|_{R_{\alpha\beta}=r_{\alpha\beta}}\equiv \tilde{g}^{\mu\nu}\sqrt{|\tilde{g}|}
\label{iLM1}
\end{equation}
It is easy to see, for instance, that this \emph{cannot} be done if $L$ has the form (\ref{NL}), while it works if the Lagrangian depends, for instance, on the squared Ricci tensor $R_{\mu\nu}R^{\mu\nu}$. For the moment, let us assume that the Lagrangian is ``Ricci--regular'', i.e.~that the inverse Legendre map $r_{\mu\nu}$ exists. Then, we apply a suitable generalization of a procedure of analytical mechanics that is completely equivalent to the usual Legendre transformation (although seldom described in current textbooks). Suppose that $L(q^{\lambda},\dot{q}^{\lambda})$ is the Lagrangian for some holonomic mechanical system on a configuration space $Q$; let $p_{\lambda}=\frac{\partial L}{\partial q^{\lambda}}$ be the Legendre map which associates to each vector in $TQ$ its conjugate momentum in $T^*Q$ and let $u^{\mu}(q^{\lambda},p_{\lambda})$ be the inverse Legendre map, such that
\begin{equation}
\left.\frac{\partial L}{\partial \dot{q}^{\lambda}}\right|_{\dot{q}^{\mu}=u^{\mu}(q^{\lambda},p_{\lambda})}\equiv p_{\lambda};\label{LM2}
\end{equation}
one can now introduce the \emph{Helmholtz Lagrangian}, which defines a holonomic system in $T^*Q$:
$$
L_H(q^{\lambda},p_{\lambda},\dot{q}^{\lambda})=p_{\mu}(\dot{q}^{\mu}-u^{\mu})+L(q^{\lambda},u^{\nu}(q^{\lambda},p_{\lambda}))
$$
(technically speaking, $L_Hdt$ is the pull-back on a curve in $T^*Q\times\mathbb{R}$ of the Poincar\'e-Cartan one-form for the Lagrangian $L$). $L_H$ is a degenerate Lagrangian, because it does not depend on $\dot{p}_{\lambda}$ and depends linearly on $\dot{q}_{\lambda}$: the resulting Euler-Lagrange equations, instead of being second-order equations (in $T^*Q$), are first order and are easily found to reproduce the dynamics of the original system: $\dfrac{\delta L_H}{\delta p_{\lambda}}=0 \Leftrightarrow \dot{q}^{\lambda}=u^{\lambda}$ and $\dfrac{\delta L_H}{\delta q^{\lambda}}=0 \Leftrightarrow \dot{p}_{\lambda}=\frac{\partial L}{\partial q^{\lambda}}$. These equations, in fact, reduce either to the Lagrange equations for $L$, taking into account (\ref{LM2}) and eliminating the conjugate momenta $p_{\mu}$, or to the Hamilton equations, upon introducing $H(q^{\lambda},p_{\lambda})=p_{\mu}u^{\mu}-L(q^{\lambda},u^{\nu}(q^{\lambda},p_{\lambda}))$ and observing that $u^{\lambda}\equiv \frac{\partial H}{\partial p_{\lambda}}$ and $\left.\frac{\partial L}{\partial q^{\lambda}}\right|_{\dot{q}^{\mu}=u^{\mu}}\equiv\frac{\partial H}{\partial q^{\lambda}}$. 

Going back to the gravitational Lagrangian (\ref{HOG}), following the same steps one finds that the Helmholtz Lagrangian
\begin{equation}
L_{H}=\tilde{g}^{\alpha\beta}(R_{\alpha\beta}-r_{\alpha\beta})\sqrt{|\tilde{g}|}+f(r_{\mu\nu},g_{\mu\nu})\sqrt{|g|}+L_{\mathrm{mat}}(j^1 \Phi,j^1 g) \label{LH1}
\end{equation}
produces Euler-Lagrange equations which are exactly equivalent to (\ref{HOGeq}). Notice that, so far, we have performed a Legendre transformation by introducing an independent ``conjugate field'', \emph{not} redefined the metric. The new tensor $\tilde{g}_{\mu\nu}$ is symmetric by definition and is generically nondegenerate, so one can rightfully regard the Lagrangian (\ref{LH1}) as describing a particular \emph{bimetric} theory. The two metrics play an unequal role, because $L_H$ depends linearly on the Ricci tensor of $g_{\mu\nu}$ and does not depend at all on derivatives of $\tilde{g}_{\mu\nu}$. However, subtracting a full divergence (which can be done without affecting the Euler-Lagrange equations), we can almost exchange the two roles. It is sufficient to use a well-known identity for the difference between the Ricci tensors of two symmetric affine connections on the same manifold:
\begin{equation}
R_{\beta\nu}-\tilde{R}_{\beta\nu}\equiv\tilde{\nabla}_{\alpha}Q^{\alpha}{}_{\beta\nu}-
\tilde{\nabla}_{\nu}Q^{\alpha}{}_{\alpha\beta}+Q^{\alpha}{}_{\alpha\sigma}Q^{\sigma}{}_{\beta\nu}-
Q^{\alpha}{}_{\beta\sigma}Q^{\sigma}{}_{\alpha\nu}\label{Riccidiff}
\end{equation}
where $\tilde{\nabla}$ denotes covariant differentiation with respect to the second connection, and the tensor $Q^{\alpha}{}_{\beta\nu}$ is the difference between the two affine connections; in our case, these are the Levi-Civita connections of the two metrics, i.e.
\begin{equation}
Q^{\alpha}{}_{\beta\nu}=\left\lbrace{}^\alpha_{\beta\nu}\right\rbrace_g-\left\lbrace{}^\alpha_{\beta\nu}\right\rbrace_{\tilde{g}}=\frac{1}{2}g^{\alpha\sigma}(\tilde{\nabla}_{\nu}g_{\sigma\beta}+\tilde{\nabla}_{\beta}g_{\sigma\nu}-\tilde{\nabla}_{\sigma}g_{\nu\beta})\label{Qmet}
\end{equation}
Once multiplied by $\sqrt{|\tilde{g}|}$, the two terms with the covariant derivatives of $Q^{\alpha}{}_{\beta\nu}$ become a full divergence. Therefore, the Helmholtz Lagrangian (\ref{LH1}) is dynamically equivalent to the following Lagrangian:
\begin{align}
L_{E}=&\ \tilde{R}_{\mu\nu}\tilde{g}^{\mu\nu}\sqrt{|\tilde{g}|}+\tilde{g}^{\alpha\beta}\left(Q^{\rho}_{\sigma\alpha}Q^{\sigma}_{\rho\beta}-Q^{\rho}_{\alpha\beta}Q^{\sigma}_{\sigma\rho}-r_{\alpha\beta}\right)\sqrt{|\tilde{g}|}\ +\nonumber\\&+f(r_{\mu\nu},g_{\mu\nu})\sqrt{|g|}+L_{\mathrm{mat}}(j^1 \Phi,j^1 g)
\end{align}
The dynamical term for $\tilde{g}_{\mu\nu}$ is now an Einstein--Hilbert term, so we immediately know that the variation of $L_{E}$ with respect to $\tilde{g}_{\mu\nu}$ produces an Einstein equation for this metric, with a complicated stress-energy tensor describing the interaction with $g_{\mu\nu}$; the full equations and their detailed analysis can be found in \cite{MS2}. However, it turns out that also the metric $g_{\mu\nu}$ obeys an Einsten equation: we know, in fact, that the variation of $L_H$ (\ref{LH1}) automatically reproduces the inverse Legendre map, i.e.~(in our case) $R_{\mu\nu}=r_{\mu\nu}(g_{\alpha\beta},\tilde{g}_{\alpha\beta})$, which can easily be recast in Einstein form (by taking its trace with $g^{\mu\nu}$). The Einstein equations for the two metrics, however, are not mutually independent: they are, in fact, two versions of the \emph{same equation}, derived from the variation with respect to $\tilde{g}_{\mu\nu}$. One could equally pass from one form to the other using (\ref{Riccidiff}), without relying on the existence of two equivalent Lagrangians.

In this picture, there is no direct interaction between $\tilde{g}_{\mu\nu}$ and the matterfields $\Phi$: this is a mere consequence of the fact that $\Phi$ has been coupled from the very beginning (\ref{HOG}) to the metric $g_{\mu\nu}$. 

In short, the introduction of the ``conjugate'' tensorfield $\tilde{g}_{\mu\nu}$ leads to an equivalent formulation of the original fourth--order equation (\ref{HOGeq}): the latter splits into two second--order equations (in much the same way as the second--order Lagrange equation, in classical mechanics, splits into a pair of first--order Hamilton equations). One of these second--order equations can be put in Einstein form, either for the original metric $g_{\mu\nu}$ or for the new tensorfield $\tilde{g}_{\mu\nu}$; only in the second case (i.e.~for the metric $\tilde{g}_{\mu\nu}$), the Einstein tensor is equated to the variational stress tensor. The latter, however, does not contain the true matterfields $\Phi$ which are instead coupled to $g_{\mu\nu}$. This situation may seem strange from the physical viewpoint, and is indeed \emph{not} physically equivalent to ordinary General Relativity with the interaction term $L_{\mathrm{mat}}$, whichever metric is considered: yet, this is nothing but an exactly equivalent representation of the dynamics described by the original Lagrangian (\ref{HOG}).

Instead of starting from (\ref{HOG}), one could first consider the \emph{vacuum} Lagrangian (deleting $L_{\mathrm{mat}}$ from (\ref{HOG})), so to keep the freedom of deciding \emph{after} the Legendre trasformation whether matter should be coupled (in the ordinary sense) either to $g_{\mu\nu}$ or to $\tilde{g}_{\mu\nu}$. This would amount to decide -- on independent grounds -- which metric should play the physical role assigned by General Relativity. Assuming that one of the two metrics defines the causal structure of spacetime, and the other, through its Levi-Civita connection, defines the free-fall geodesics, would be physically untenable: generically, the two metric are not conformal to each other, and the inertial structure of space-time would then become incompatible with the causal structure (in the sense of Ehlers--Pirani--Schild). Thus, either $g_{\mu\nu}$ and $\tilde{g}_{\mu\nu}$ should be regarded as providing both causal and inertial structure.
The other tensorfield, then, could be assimilated to a gravitating massive field, and might be conveniently split into a scalar field (the trace $g_{\mu\nu}\tilde{g}^{\mu\nu}$) and a traceless symmetric tensor, corresponding to a spin-0 and a spin-2 field respectively (see \cite{MS2} for a complete discussion). The dynamical term for the spin-2 field, in both cases, has unphysical signature (still, it is the only known consistent model of a spin-2 field interacting with gravity). 

Now, let us revert to the restricted case (\ref{NL}), which enjoyed a much larger popularity in cosmology. As we have already remarked, from the viewpoint of the Legendre transformation this is \emph{not} a particular case of the setup described in the previous part of this section. The Lagrangian (\ref{NL}), in fact, is not ``Ricci-regular'', meaning that the Legendre map (\ref{LM1}) cannot be inverted to re-express the components of the Ricci tensor as functions of $g_{\mu\nu}$ and $\tilde{g}_{\mu\nu}$. This is simply because the r.h.s.~of (\ref{LM1}) does not contain now the full Ricci tensor, but merely its trace. According to the mathematical framework which is explained in \cite{MFF90}, since the Lagrangian (\ref{NL}) depends only on the curvature scalar $R$, then it is the (scalar) conjugate field to $R$ which should be introduced by a suitable Legendre map:
\begin{equation}
p=\frac{1}{\sqrt{|g|}}\frac{\partial L}{\partial R}=f'(R).\label{p1}
\end{equation}
Then, whenever $f''(R)\neq 0$ (we may say that the Lagrangian is ``R-regular''  in this case), one can invert (\ref{p1}) and get a function $r(p)$ (inverse Legendre map) such that
\begin{equation}
\left.f'(R)\right|_{R=r(p)}.\equiv p\label{ip1}
\end{equation}
The Helmholtz Lagrangian for this case is then
\begin{equation}
L_{H}=p(R-r)\sqrt{-g}+f(r)\sqrt{|g|}+L_{\mathrm{mat}}(j^1\Phi,j^1g).\label{LH}
\end{equation}
Instead of a bimetric theory, the Legendre transformation now led us to a scalar-tensor theory. The scalar field $p$, however, has no separate dynamical term. As is well known, at this point one can perform a conformal rescaling of the metric (whenever $p>0$), so that the coupling of $p$ with the Ricci tensor is replaced by an ordinary quadratic first-order term. Setting
\begin{equation}
\tilde{g}_{\mu\nu}=pg_{\mu\nu}\label{cr}
\end{equation}
and subtracting a full divergence, the Helmholtz Lagrangian transforms into
\begin{align}
L_{E}=&\ \tilde{R}\sqrt{|\tilde{g}|}+\frac{1}{p^{2}}\left(-\frac{3}{2}\tilde{g}^{\mu\nu}\partial_{\mu} p\, \partial_{\nu} p+
f(r)-p\cdot r(p)\right)\sqrt{|\tilde{g}|}\ +\nonumber\\ &+L_{\mathrm{mat}}(j^1\Phi,j^1g),
\end{align}
which can be put in more familiar form by redefining also the scalar field by  $p=e^{\sqrt{\frac{3}{2}}\varphi}$:
\begin{equation}
L_{E}=\left(\tilde{R}-\tilde{g}^{\mu\nu}\varphi_{,\mu}\varphi_{,\nu}-2V(\varphi)\right)\sqrt{|\tilde{g}|}+L_{\mathrm{mat}}(j^1\Phi,j^1g).
\end{equation}
The last term can, in principle, rewritten so that matter fields $\Phi$ appear to interact with $\varphi$ and $\tilde{g}_{\mu\nu}$. The equation for the gravitational field can be written as an Einstein equation for either the ``Jordan'' metric $g_{\mu\nu}$ or the rescaled ``Einstein'' metric $\tilde{g}_{\mu\nu}$. Only in the second case, however, the Einstein tensor is equated to a variational stress--energy tensor, which is covariantly conserved along the connection $\lbrace{}^\alpha_{\beta\nu}\rbrace_{\tilde{g}}$. The r.h.s.~of the Einstein equation for $g_{\mu\nu}$ can be viewed as an ``effective energy--momentum tensor'', and is covariantly conserved along the connection $\lbrace{}^\alpha_{\beta\nu}\rbrace_g$, but does not coincide with the variational derivative of the interaction terms of the Lagrangian. Furthermore, whenever $f''(R)>0$ at $R=0$,  \emph{in the Einstein frame} the stress--energy tensor for the scalar field fulfills the dominant energy condition \cite{MS1}: this provides a possible criterion (that we mentioned in Section 1) to consider the Einstein metric $\tilde{g}_{\mu\nu}$ to be the physical one. If this criterion is adopted, then one should ensure that the coupling with matter fields $\Phi$ becomes the physically appropriate one when expressed in terms of the Einstein metric.

It is striking to remark that the rescaled metric $\tilde{g}_{\mu\nu}$, even in this case, meets the Kijowski prescription: since for $f(R)$ one has \mbox{$\frac{\partial f}{\partial R_{\mu\nu}}= f'(R)g^{\mu\nu}$}, we see that (\ref{LM1}) holds true. The idea of defining a metric as the conjugate momentum to the Ricci tensor is somehow independent of which Legendre map is actually invertible (and should therefore be used to perform a Legendre transformation): this fact is at the core of the ``universality of the Einstein equation" repeatedly advocated by J.~Kijowski himself and by many other authors after him. 

On the other hand, it is worthwhile to stress again the differences between the dynamics generated by Ricci-regular Lagrangians (\ref{HOG}) and by R-regular Lagrangians (\ref{NL}). In the Ricci-regular case, the model is equivalent to a bimetric theory, while in the R-regular case the gravitational degrees of freedom are represented by a metric and a scalar field. In the first case, the two metrics coexist as independent dynamical variables in the second-order formulation of the model. In the $f(R)$ case, instead, the Jordan metric and the Einstein metric are conformally related, and only either of them can appear in the field equations (jointly with the scalar field); the Einstein metric does not pop up from the Legendre transformation, but from the subsequent (and independent) conformal rescaling.

\section{Metric--affine (Palatini) theories}

In the previous section we have basically rephrased the results presented in \cite{NGL} (taking into account the further insight given by the more detailed analysis in \cite{MS1} and \cite{MS2}). Let us now consider the metric--affine version of these models, adopting the same approach. Here, the affine connection $\Gamma^{\lambda}_{\mu\nu}$ is not assumed to be metric, and becomes and independent dynamical field. \smallskip

In a lecture given in Turin a few years ago, Mauro Francaviglia quoted a celebrated passage that can be found in Galileo Galilei's \emph{Saggiatore}:

\emph{``Philosophy is written in this grand book, which stands continually open before our eyes (I say the Universe), but cannot be understood without first learning to comprehend the language and know the characters in which it is written. It is written in mathematical language, and its characters are triangles, circles and other geometric figures, without which it is impossible for humans to understand a word; without these, one is wandering in a dark labyrinth.''}

Mauro remarked that Galileo's reference to ``triangles'', geometric figures whose definition implies the notion of \emph{straight line}, i.e.~geodesic, and to ``circles'' -- figures which encapsulate the \emph{metric} structure of space -- may sound as a striking anticipation of the modern view of the structure of the Universe. He confronted this idea with the Palatini framework (which is actually due to Einstein), where the geodesic and the causal structures are kept independent in principle, and become related only by the dynamical equations. We had no opportunity of learning from Mauro whether this fascinating and daring parallel to Galileo's text was due to himself or to another author, but it is definitely representative of Mauro's penetrating view on research in mathematical physics.
\smallskip

We shall denote by $\mathcal{R}_{\mu\nu}$ the Ricci tensor of the connection $\Gamma^{\lambda}_{\mu\nu}$, and by $\mathcal{R}$ its trace with the metric, $\mathcal{R}=\mathcal{R}_{\alpha\beta}g^{\alpha\beta}$. We shall restrict to what is commonly considered the ``Palatini setup'' (in contrast to the most general metric--affine setup), by the following three relevant assumptions:
\begin{description}
\item[(A)] $\Gamma^{\lambda}_{\mu\nu}$ is symmetric: the torsion of the connection (i.e.~its antisymmetric part, which does not enter the geodesic equation) would have a non--gravitational physical interpretation, which is beyond the scope of our discussion;
\item[(B)] the Lagrangian depends only on the symmetric part of the Ricci tensor; the possible role of the antisymmetric part $\mathcal{R}_{[\mu\nu]}$ (which vanishes identically if the connection is metric) has been studied many years ago \cite{FK}  and is known to produce the appearance of a spin-1, massive (Proca) field in the model, of doubtful physical interpretation;
\item[(C)] the matterfields $\Phi$ interact only with the metric: the matter Lagrangian, which in principle could be expected to be of the form  $L_{\mathrm{mat}}(j^1 \Phi,g,\Gamma)$, is instead $L_{\mathrm{mat}}(j^1 \Phi,j^1 g)$. In other terms, possible covariant derivatives of the matterfields are defined using the Levi-Civita connection of $g$ instead of $\Gamma$. A sound physical motivation for this assumption has never been given, to our knowledge, but it turns out  that a direct coupling of matter with $\Gamma$, as will become clear in the next paragraphs, would radically change the situation and make it much more distant from General Relativity: hence, some authors purposely reserve the name "Palatini theories" only for models where the matter is not coupled to the independent connection $\Gamma$ \cite{Cap}. 
\end{description}

\noindent With the above assumptions, the Palatini counterpart of Lagrangian (\ref{HOG}) becomes
\begin{equation}
L=f(\mathcal{R}_{(\mu\nu)},g_{\mu\nu})\sqrt{|g|}+L_{\mathrm{mat}}(j^1 \Phi,j^1 g),\label{PalRic}
\end{equation}
and the counterpart of (\ref{NL}) is
\begin{equation}
L=f(\mathcal{R})\sqrt{|g|}+L_{\mathrm{mat}}(j^1 \Phi,j^1 g)\label{PalR}.
\end{equation}
Contrary to the previous Section, for the reader's convenience we shall first consider the $f(\mathcal{R})$ case (\ref{PalR}), since it is much more popular in the current literature and the resulting picture will be somehow easier to interpret. 

In the vacuum case ($L_{\mathrm{mat}}\equiv 0$), the Euler-Lagrange equations read
\begin{align}
&\bar{\nabla}_{\lambda}\left(f'(\mathcal{R})g^{\mu\nu}\sqrt{|g|}\right) = 0  \label{met}\\ 
&f'(\mathcal{R})\mathcal{R}_{\mu\nu}-\frac{1}{2}f(\mathcal{R})g_{\mu\nu} = 0,\label{ME0}
\end{align}
where $\bar{\nabla}$ denotes covariant derivation with the connection $\Gamma$. As is well known, taking the trace of the second equation (\ref{ME0}) with $g^{\mu\nu}$ one gets an algebraic equation for $\mathcal{R}$, 
$$
f'(\tilde{R})\tilde{R}-2f(\tilde{R})=0\quad\Rightarrow\quad \tilde{R}=c_k
$$
(in dimension four) where the roots $c_k$ (and their number) only depend on the function $f$ in the Lagrangian. $\mathcal{R}$ being constant, $f'(\mathcal{R}$ should be constant, too, and equation (\ref{met}) implies that $\Gamma^{\lambda}_{\mu\nu}$ (under the assumption (A) above) should coincide with the Levi-Civita connection $\left\lbrace{}^\alpha_{\beta\nu}\right\rbrace_g$. Therefore, $\mathcal{R}_{\mu\nu}$ coincides with $R_{\mu\nu}$, the Ricci tensor of the metric. Equation (\ref{ME0}) then becomes an Einstein equation with cosmological constant $\Lambda$ (whose value depends on the function $f$ and on the particular root $c_k$ considered: $\Lambda_k=\frac{1}{2}\left(\frac{f(c_k)}{f'(c_k)}-c_k\right)$). Hence, the set of solutions of a vacuum Palatini $f(\mathcal{R})$ is exactly the union of the solutions of vacuum General Relativity over the set of possible values $\Lambda_k$ of the cosmological constant \cite{FFV}.

Let us now switch on matter interaction. Equation (\ref{met}), coming from the variation of  (\ref{PalR}) with respect to the connection, is not affected (it is here that assumption (C) plays a crucial role); (\ref{ME0}), instead, becomes
\begin{equation}
f'(\mathcal{R})\mathcal{R}_{\mu\nu}-\frac{1}{2}f(\mathcal{R})g_{\mu\nu} =\kappa T_{\mu\nu}.\label{ME1}
\end{equation}
Its trace is again an algebraic equation for the scalar $\mathcal{R}$, but now the r.h.s. is, in general, a function of $j^1\Phi$ (and possibly of $j^2 g$). Hence, the roots \mbox{$\mathcal{R}=c_k(T)$} will be non--constant functions in spacetime, through the trace $T$ of the stress-energy tensor. Therefore, $f'(\mathcal{R})$ will no longer be constant: equation (\ref{met}), then, implies that $\Gamma^{\lambda}_{\mu\nu}$ coincides with the Levi-Civita connection $\lbrace{}^\alpha_{\beta\nu}\rbrace_{\tilde{g}}$ for a different metric tensor, namely the conformally rescaled metric fulfilling
\begin{equation}
\tilde{g}^{\mu\nu}\sqrt{|\tilde{g}|}=f'(c_k(T))g^{\mu\nu}\sqrt{|g|}.
\end{equation}
Let us reinterpret these well--known facts, by applying a Legendre transformation: analogously to (\ref{p1}) we set
\begin{equation}
p=\frac{1}{\sqrt{|g|}}\frac{\partial L}{\partial \mathcal{R}}=f'(\mathcal{R}),\label{p2}
\end{equation}
we introduce the inverse Legendre map $r=r(p)$ and we obtain the (Palatini) Helmholtz Lagrangian
\begin{equation}
L_{H}=p(\mathcal{R}-r)\sqrt{|g|}+f(r)\sqrt{-g}+L_{\mathrm{mat}}(j^1 \Phi,j^1 g).\label{LHP}
\end{equation}
The difference, with respect to (\ref{LH}), is that the dynamical fields are now the connection $\Gamma$, the scalar field $p$ and the metric $g$. The variations with respect to these three fields, in this order, yield the following equations, which are manifestly equivalent to (\ref{met}) and (\ref{ME1}):
\begin{align}
& \bar{\nabla}_{\lambda}\left(p g^{\mu\nu}\sqrt{|g|}\right)=0\label{met2} \\[6pt]
& \mathcal{R}_{\alpha\beta}g^{\alpha\beta}=r(p)\label{uf1}\\ 
& p\mathcal{R}_{\mu\nu}-\frac{1}{2}f(r(p))g_{\mu\nu}=T_{\mu\nu}.\label{uf2}
\end{align}
We know that eq.~(\ref{met2}) is equivalent to 
\begin{equation}
\Gamma^\lambda_{\mu\nu}=\left\lbrace{}^\lambda_{\mu\nu}\right\rbrace_{\tilde{g}},\quad\text{where}\quad \tilde{g}_{\mu\nu}=pg_{\mu\nu};\label{vin}
\end{equation}
this means that the dynamics can be completely described by one metric and one scalar field. One has, as a matter of fact, two possibilities: either one chooses to represent the dynamics in terms of the scalar field $p$ and the metric $g_{\mu\nu}$, or one chooses the pair $(p, \tilde{g}_{\mu\nu})$. The steps are as follows: first, one observes that the lower--order relations (\ref{vin}), being \emph{identically satisfied for all solutions} of the field equations, can be safely plugged into in the Lagrangian itself. This amounts to substituting the Ricci tensor $\mathcal{R}_{\mu\nu}$ with the Ricci tensor of $\tilde{g}_{\mu\nu}$, obtaining
\begin{equation}
L_{H}=p(\tilde{R}_{\mu\nu}g^{\mu\nu}-r)\sqrt{|g|}+f(r)\sqrt{-g}+L_{\mathrm{mat}}(j^1 \Phi,j^1 g).\label{LHP2}
\end{equation}
On account of the relation between the two metrics (\ref{vin}), one can immediately get rid of the metric $g_{\mu\nu}$, and in dimension four the Lagrangian (in what can be called the Einstein frame) becomes
\begin{equation}
L_{E}=\tilde{R}\sqrt{|\tilde{g}|}+\left(p^{-2}f(r)-p^{-1}r\right)\sqrt{|\tilde{g}|}+L_{\mathrm{mat}}(j^1\Phi,j^1\tilde{g},j^1p).\label{LEH2}
\end{equation}
To keep the Jordan--frame metric, instead, one can exploit once more the identity for the difference of the two Ricci tensors; in the case of two con\-for\-mal\-ly--related metrics, it is known that the difference of the two Levi--Civita connections can be expressed in terms of the first derivatives of the conformal factor. The resulting Jordan--frame Lagrangian (\cite{Cap}, \cite{FF}) is 
\begin{equation}
L_{J}=pR\sqrt{|g|}+\left(\frac{3}{2p}g^{\mu\nu}\partial_\mu p\, \partial_\nu p-p\cdot r+f(r)\right)\sqrt{|g|}+L_{\mathrm{mat}}(j^1\Phi,j^1g).
\end{equation}
For the reader's convenience, let us put alongside the Legendre--transformed versions of purely metric and Palatini $f(R)$ theories:
\medskip\par
\noindent\fbox{ 
\addtolength{\linewidth}{-20pt}
\begin{minipage}{\linewidth}
\medskip\par
\begin{center}
{\bf Purely metric:}
\end{center}
\begin{equation*}
L=f(R)\sqrt{|g|}+L_{\mathrm{mat}}(j^1 \Phi,j^1 g)
\end{equation*}\smallskip

\noindent Equivalent scalar--tensor Lagrangian (Jordan frame):
\begin{equation*}
L_{J}=pR\sqrt{|g|}+\left(f(r)-p\cdot r)\right)\sqrt{|g|}+L_{\mathrm{mat}}(j^1\Phi,j^1g)
\end{equation*}
Einstein--frame Lagrangian:
\begin{equation*}
L_{E}= \tilde{R}\sqrt{|\tilde{g}|}+\frac{1}{p^{2}}\left(-\frac{3}{2}\tilde{g}^{\mu\nu}\partial_{\mu} p\, \partial_{\nu} p+
f(r)-p\cdot r\right)\sqrt{|\tilde{g}|}+L_{\mathrm{mat}}(j^1\Phi,j^1\tilde{g},j^1p)
\end{equation*}
\hrule
\begin{center}
{\bf Palatini:}
\end{center}
\begin{equation*}
L=f(\mathcal{R})\sqrt{|g|}+L_{\mathrm{mat}}(j^1 \Phi,j^1 g)
\end{equation*}\smallskip

\noindent
Equivalent scalar--tensor Lagrangian (Jordan frame):
\begin{equation*}
L_{J}=pR\sqrt{|g|}+\left(\frac{3}{2p}g^{\mu\nu}\partial_\mu p\, \partial_\nu p+f(r)-p\cdot r\right)\sqrt{|g|}+L_{\mathrm{mat}}(j^1\Phi,j^1g).
\end{equation*}
Einstein--frame Lagrangian:
\begin{equation*}
L_{E}=\tilde{R}\sqrt{|\tilde{g}|}+\frac{1}{p^{2}}\left(f(r)-p\cdot r\right)\sqrt{|\tilde{g}|}+L_{\mathrm{mat}}(j^1\Phi,j^1\tilde{g},j^1p)\end{equation*}
\smallskip
\end{minipage}
}
\par\medskip\noindent
In both cases, $r=r(p)$ is the map such that $f'(x)|_{x=r(p)}\equiv p$. Notice that the coupling between $\Phi$ and the scalar field $p$ in the matter Lagrangian in the Einstein frame is not unavoidable. In some relevant cases (e.g.~scalar fields, electromagnetic field, cosmic dust), the matter Lagrangian contains only ordinary derivatives of $\Phi$, not covariant ones, so it would be $L_{\mathrm{mat}}(j^1\Phi,\tilde{g},p)$; then, it may be possible to devise a suitable non--minimal coupling in the original Lagrangian (which should then be more appropriately referred to as a $f(R,\Phi)$ Lagrangian), so that the dependence of $L_{\mathrm{mat}}$ on $p$ disappears in the Einstein frame picture (various examples are given in \cite{MS1}).

Direct comparison of the boxed expressions shows that the difference between the purely metric models and the corresponding Palatini models entirely consists in the fact that the dynamical term for the scalar field, which in the purely metric setup is absent in the Jordan frame and appears in the Einstein frame, for the Palatini models is instead found (with opposite sign) in the Jordan-frame scalar-tensor Lagrangian, but clears away in the Einstein frame. 

This difference has, indeed, relevant consequences. For instance, in vacuum the scalar field $p$ has still a nontrivial dynamics in the purely metric setup, while in the Palatini setup $p$ is forced to be constant in spacetime and can assume a set of possible values being the roots of the equation $p\cdot r(p)-2f(r(p))=0$, which is the trace of eq.~(\ref{uf2}) combined with (\ref{uf1}). It is easy to see that these are nothing but the particular solutions of the purely metric model for which $p$ is constant. Thus, we see that in vacuum the set of solutions for the Palatini model are a \emph{proper subset} of the solutions on the purely metric model (this fact was first observed in \cite{mio}). In the presence of matter, this is no longer true. The field $p$ cannot be constant unless $T$, the trace of the matter stress-energy tensor, is constant as well: in general the dynamical term for $p$, which makes the difference between the two models, cannot vanish. It is still true, however, that in the purely metric theory $p$ behaves (in the Einstein frame) as a true independent dynamical scalar field, while in the Palatini framework it becomes a mere function of the metric and of the matter/energy distribution, without independent propagation. 

Let us eventually turn to the Ricci-regular case. For brevity, in the sequel we drop the matter interaction term and consider only the vacuum Lagrangian. As usual, we introduce the Legendre map for the Lagrangian (\ref{PalRic}), and produce a symmetric tensor from the conjugate momentum (which is a tensor density):
\begin{equation}
\tilde{g}^{\mu\nu}\sqrt{|\tilde{g}|}=\frac{\partial L}{\partial \mathcal{R}_{(\mu\nu)}}=\frac{\partial f}{\partial\mathcal{R_{(\mu\nu)}}}\sqrt{|g|}.
\label{LM3}
\end{equation}
Setting $r_{\mu\nu}=r_{\mu\nu}(g_{\alpha\beta},\tilde{g}_{\alpha\beta})$ to be the symmetric tensor-valued function being the inverse of the Legendre map (notice that by construction $r_{\mu\nu}$ does not depend on the connection $\Gamma$), the Helmholtz Lagrangian becomes
\begin{equation}
L_{H}=\tilde{g}^{\alpha\beta}(\mathcal{R}_{(\alpha\beta)}-r_{\alpha\beta})\sqrt{|\tilde{g}|}+f(r_{\mu\nu},g_{\mu\nu})\sqrt{|g|}.
\end{equation}
Taking the variation of the original Lagrangian with respect to the connection $\Gamma^{\lambda}_{\mu\nu}$, one finds the equation
\begin{equation}
\bar{\nabla}_{\lambda}\left(\dfrac{\partial f}{\partial\mathcal{R}_{(\mu\nu)}}\sqrt{-g}\right)=0,
\end{equation}
which in the Legendre--transformed picture (i.e., taking the corresponding variation of $L_H$) becomes
\begin{equation}
\bar{\nabla}_{\lambda}\left(\tilde{g}^{\mu\nu}\sqrt{|\tilde{g}|}\right)=0,
\end{equation}
which means that the (symmetric) connection $\Gamma^{\lambda}_{\mu\nu}$ should be the Levi--Civita connection of the metric $\tilde{g}^{\mu\nu}$. This holds true even in the presence of matter, \emph{provided} the matter Lagrangian does not contain $\Gamma^{\lambda}_{\mu\nu}$ (assumption (C) above). Hence, such a metric-affine theory is equivalent to a bimetric theory where the (independent) dynamical fields are $g^{\mu\nu}$ and $\tilde{g}^{\mu\nu}$, with the Lagrangian
\begin{equation}
L_{E}=\tilde{g}^{\alpha\beta}\tilde{R}_{\alpha\beta}+\left(f(r_{\mu\nu},g_{\mu\nu})\sqrt{|g|}-r_{\alpha\beta}\tilde{g}^{\alpha\beta}\sqrt{|\tilde{g}|}\right).\label{uf3}
\end{equation}
One therefore finds directly an Einstein--frame Lagrangian, in contrast to (\ref{LH1}). Here, it is the Ricci tensor of the metric $\tilde{g}_{\mu\nu}$ which enters the Helmholtz Lagrangian, and therefore the dynamical term for $g_{\mu\nu}$, which in the purely metric setup appears in the Einstein frame after replacing $R_{\alpha\beta}$ with $\tilde{R}_{\alpha\beta}$, is absent from (\ref{uf3}). Once again, we summarize the outcome of the discussion in the following box:
\medskip\par
\noindent\fbox{ 
\addtolength{\linewidth}{-20pt}
\begin{minipage}{\linewidth}
\medskip\par
\begin{center}
{\bf Purely metric:}
\end{center}
\begin{equation*}
L=f(R_{\mu\nu},g_{\mu\nu})\sqrt{|g|}
\end{equation*}\smallskip

\noindent Equivalent bimetric Lagrangian (Jordan frame):
\begin{equation*}
L_{J}=\tilde{g}^{\alpha\beta}R_{\alpha\beta}\sqrt{|\tilde{g}|}+\left(f(r_{\mu\nu},g_{\mu\nu})\sqrt{|g|}-r_{\alpha\beta}\tilde{g}^{\alpha\beta}\sqrt{|\tilde{g}|}\right)
\end{equation*}
Einstein--frame Lagrangian:
\begin{align*}
L_{E}=& \tilde{g}^{\alpha\beta}\tilde{R}_{\alpha\beta}\sqrt{|\tilde{g}|}+\left(f(r_{\mu\nu},g_{\mu\nu})\sqrt{|g|}-r_{\alpha\beta}\tilde{g}^{\alpha\beta}\sqrt{|\tilde{g}|}\right)+\\&+\tilde{g}^{\alpha\beta}\left(Q^{\rho}_{\sigma\alpha}Q^{\sigma}_{\rho\beta}-Q^{\rho}_{\alpha\beta}Q^{\sigma}_{\sigma\rho}\right)\sqrt{|\tilde{g}|}
\end{align*}
\hrule
\begin{center}
{\bf Palatini:}
\end{center}
\begin{equation*}
L=f(\mathcal{R}_{(\mu\nu)},g_{\mu\nu})\sqrt{|g|}
\end{equation*}\smallskip

\noindent
Equivalent bimetric Lagrangian (Einstein frame):
\begin{equation*}
L_{E}=\tilde{g}^{\alpha\beta}\tilde{R}_{\alpha\beta}\sqrt{|\tilde{g}|}+\left(f(r_{\mu\nu},g_{\mu\nu})\sqrt{|g|}-r_{\alpha\beta}\tilde{g}^{\alpha\beta}\sqrt{|\tilde{g}|}\right)\end{equation*}
\smallskip
\end{minipage}
}
\par\medskip\noindent
Again, we see that the difference lies in the dynamical term for the metric $g_{\mu\nu}$, which appears in the Einstein frame Lagrangian for the purely metric framework, and does not occur in the Einstein frame Lagrangian for the Palatini model. This term can be written as a quadratic term in the covariant derivatives $\tilde{\nabla}_{\lambda}g_{\mu\nu}$, and can be further decomposed into a standard dynamical term for a scalar field and a dynamical term for a spin-two field. 

In the Palatini case, instead, the tensor $g_{\mu\nu}$ does not propagate independently: its configuration depends on $\tilde{g}_{\mu\nu}$ and on the possible coupling with matter in the original Lagrangian. There are particular cases, e.g.~Palatini Lagrangians depending only on the square of the Ricci tensor, $f(\mathcal{R}_{\alpha\beta}\mathcal{R}^{\alpha\beta})$ where (in vacuum) the equations completely reduce to Einstein equations for the original metric $g_{\mu\nu}$ alone \cite{Bor}. Otherwise, the two metrics are not even conformally related, and in general one should not expect that the causal structure defined by $g_{\mu\nu}$ and the geodesic worldlines associated to $\tilde{g}_{\mu\nu}$ (and therefore to the original connection $\Gamma^{\lambda}_{\mu\nu}$) can be compatible in the EPS sense. This somehow undermines the physical consistency of generic Ricci-regular Palatini models, unless one is willing to assume that $\tilde{g}_{\mu\nu}$ is the physical metric (or, alternatively, that the inertial structure is defined by the Levi-Civita connection of $g_{\mu\nu}$, rather than by $\Gamma^{\lambda}_{\mu\nu}$).

\section*{Acknowledgments}

The author wishes to thank Marco Ferraris, Leszek \mbox{Soko\l{}owski} and Lorenzo \mbox{Fatibene} for innumerable fruitful discussions in the past years; Jerzy \mbox{Kijowski}, Andrzej \mbox{Borowiec} and Demeter Krupka for sound remarks during the workshop. 

\noindent This article has been written as a tribute to the memory of Mauro \mbox{Francaviglia}, who initiated the author into scientific research.

\end{document}